\documentclass{article}
\usepackage{latexsym}
\usepackage{braket}
\usepackage{mathrsfs}
\usepackage{mathtools}
\usepackage{color}
\usepackage{xcolor}
\usepackage{subcaption}
\usepackage{enumitem}
\usepackage[toc,page]{appendix}
\usepackage{cite}
\usepackage[english]{babel}
\usepackage{babelbib}
\usepackage{multirow}
\usepackage{tabularx}
\usepackage[T1]{fontenc}
\usepackage{makecell}%To keep spacing of text in tables
\setcellgapes{4pt}%parameter for the spacing
\usepackage[a4paper, total={6in, 8in}]{geometry}

\usepackage[thinlines]{easytable}

\usepackage{colortbl}
\def\bal#1\eal{\begin{align}#1\end{align}}
\usepackage{color,soul}
\newcommand{\bsub}{\begin{subequations}}
\newcommand{\esub}{\end{subequations}}
\def\bal#1\eal{\begin{align}#1\end{align}}

\newcommand{\gra}{{\alpha}} \newcommand{\grb}{{\beta}} \newcommand{\grg}{{\gamma}} \newcommand{\grd}{{\delta}}
\newcommand{\gre}{{\epsilon}}   \newcommand{\gru}{{\theta}}
 \newcommand{\grk}{{\kappa}} \newcommand{\grl}{{\lambda}} \newcommand{\grm}{{\mu}}
\newcommand{\grn}{{\nu}} \newcommand{\grj}{{\xi}}  \newcommand{\grp}{{\pi}}
\newcommand{\grr}{{\rho}} \newcommand{\grs}{{\sigma}} \newcommand{\grt}{{\tau}} 
\newcommand{\grf}{{\phi}} \newcommand{\grx}{{\chi}} \newcommand{\grc}{{\psi}} 
 \newcommand{\grB}{{\rm B}}  
   
  \newcommand{\grL}{{\Lambda}} 
   \newcommand{\grP}{{\Pi}}
   
\newcommand{\grF}{{\Phi}}  \newcommand{\grC}{{\Psi}} \newcommand{\grV}{{\Omega}}

\begin{document}

\title{\textbf{``Time''-covariant Schr\"{o}dinger equation and the canonical quantization of the Reissner-Nordstr\"{o}m black hole}}
\author{ \textbf{T. Pailas}$^a$\thanks{teopailas879@hotmail.com}\\
\normalsize
{$^a$\it Department of Nuclear and Particle Physics, Faculty of Physics,}\\
\normalsize{\it National and Kapodistrian University of Athens, Athens 15784, Greece}}

\maketitle

\abstract{A ``time''-covariant Schr\"{o}dinger equation is defined for the minisuperspace model of the Reissner-Nordstr\"{o}m (RN) black hole, as a ``hybrid''  between the ``intrinsic time'' Schr\"{o}dinger and Wheeler-DeWitt(WDW) equations. To do so, a reduced, regular and ``time(r)''-dependent Hamiltonian density was constructed, without ``breaking'' the re-parametrization covariance $r\rightarrow f(\tilde{r})$. As a result, evolution of states with respect to the parameter $r$ and probabilistic interpretation of the resulting quantum description is possible, while quantum schemes for different gauge choices are equivalent by construction. The solutions are found for a Dirac's delta and a Gaussian initial states. A geometrical interpretation of the wavefunctions is presented via Bohm analysis. Alongside, a criterion is presented to adjudicate which, between two singular spacetimes is ``more'' or ``less'' singular. Two ways to adjudicate about the existence of singularities are compared (vanishing of the probability density at the classical singularity and semi-classical spacetime singularity). Finally, an equivalence of the reduced equations with these of a 3D electromagnetic pp-wave spacetime is revealed.}

\newpage

\section{Introduction}
The quantization of gravity possesses a fundamental place in the realm of theoretical physics. As it was argued by Claus Kiefer \cite{Kiefer:2013jqa}, there are mainly three arguments that motivate such a quest: the first motivation is based more on philosophical reasons and the idea that all fundamental interactions should be unified at some energy scale. Hence, if a coherent quantum theory exists that describes those fundamental interactions, gravity should be included as well. The second motivation is related to the appearance of singularities in the classical theory of gravity, at it is described by general relativity. The hope, is that a quantum theory of gravity will be free of such pathological situations. Finally, the third motivation is related to the so called ``problem of time'' as can be found in \cite{Kuchar:1991qf, PhysRevD.40.2598,Halliwell:1994wq,Isham:1992ms,Barvinsky:1993jf,Anderson:2010xm}. The main reason that such a problem arises is that in the conventional quantum theory, time is an external parameter, while in the general theory of relativity, there is no preferred or fundamental notion of time. This is due to the diffeomorphism invariance of the theory. Furthermore, this invariance results in a singular Lagrangian and Hamiltonian density respectively which implies constraint equations in addition to the dynamical. The relation between constraint equations and diffeomorphisms can be found in \cite{Pons:2009cz} as well as in \cite{Barbour:2008bw}.

The appearance of constraints results in two main approaches in the quantization procedure: the first is to solve the constraints before quantization, while the second is to impose the constraints at the quantum level, as restrictions upon the states. One of the main paths to deal with the above approaches is the so called ``canonical quantization'', where the canonical variables that describe the gravitational field are promoted to operators and the Poisson brackets are replaced by operator commutators. Some of the original work in the direction of the Hamiltonian formalism of gravity is given in \cite{10.2307/100496,Dirac:1958sc,PhysRev.116.1322,PhysRev.116.1324,PhysRev.116.1325,PhysRev.116.1326,PhysRev.116.1327,PhysRev.116.1328,PhysRev.116.1329,PhysRev.116.13210}.

Related to the second procedure, the most famous result is the Wheeler-DeWitt equation \cite{Wheeler:1964qna, Wheeler:1988zr, DeWitt:1962cg, PhysRev.160.1113, PhysRev.162.1195, PhysRev.162.1239} which results from the quadratic constraint, or in other words from the variation of the gravitational action with respect to the Lapse function, in the canonical formalism. This equation contains no external parameter in which the term ``time'' could be assigned. In fact, due to its hyperbolic nature, the ``time'' parameter is encoded in the degrees of freedom that appear, hence the name ``intrinsic'' time. Thus, additional hypothesis are needed in order to define a ``time'' and hence arrive in some sort of a Schr\"{o}dinger equation. On of the problems arising due to the hyperbolic nature is the difficulty in defining a proper Hilbert space as in the conventional quantum theory.

When it comes to the first procedure, the usual approach is to introduce quantities $\grx^{A}$, which will be functional of the degrees of freedom and impose a suitable gauge choice. To this end, a ``true'' Hamiltonian density $h_{A}$ can be constructed, which depends only on the true gravitational degrees of freedom. The constraint equation has the form 
\begin{align}
P_{A}+h_{A}=0,\label{sdsfd}
\end{align}
where $P_{A}$ are the conjugate momenta of the introduced quantities $\grx^{A}$. A Schr\"{o}dinger type equation can then be realized from the above equation, with the identification $P_{A}=i h \frac{\grd}{\grd \grx^{A}}$
\begin{align}
i h \frac{\grd \grC}{\grd \grx^{A}}=\hat{h}_{A}\grC.\label{sfdgfv}
\end{align}
An extensive analysis of this approach can be found in \cite{Isham:1992ms, PhysRev.126.1864, Kuchar:1971xm}. Now, since in the constraint, the momenta appear quadratically, in many cases it is very difficult to obtain a closed form density $h_{A}$. This is one of the reasons that many authors have assumed the existence of matter in order to define a standard of time \cite{Brown:1994py, Alexander:2012tq, Bojowald:2010xp, Rovelli:1989jn}. In this case, a Hilbert space can be defined and alongside a probabilistic interpretation could be provided \cite{Isham:1992ms}. As it was argued by Barvinsky \cite{Barvinsky:1993jf}, choosing a gauge breaks the gauge invariance of the theory and hence, for each different gauge (classically equivalent systems) one has different quantum theories, which must be proven to be equivalent. Another problem, it is that the Hamiltonian density $h_{A}$ is the square root of the true degrees of freedom and the ``time'' variables, which implies that in many cases it will remain real only for a specific domain of the ``time'' variable. Furthermore, since it is a square root operator it is known that it can be handled only if the quantity whose root is being taken, is self-adjoint and positive \cite{Isham:1992ms}. Note also that $\hat{h}_{A}$ are ``time'' dependent which implies that we cannot reproduce the Wheeler-DeWitt equation from \eqref{sfdgfv}. This is not something to worry about, it simply reveals the different quantization approaches.  An interesting new approach to the problem of time can be found in \cite{Hoehn:2019owq}.

These interesting questions are very difficult to be tackled in the full theory of gravity. In order to acquire some insight, people have worked on simplified models such as minisuperspace Lagrangians, where the isometries of spacetime are enough, in order to render the system from infinite to a finite dimensional. In cosmology, such examples are the FLRW geometries, the Bianchi Types and the Kantowski-Sachs metric \cite{Ryan}. Additionally, there are the point-like sources such as Schwarzschild (S) \cite{Schwarzschild:1916uq} and Reissner-Nordstr\"{o}m (RN) \cite{1916AnP...355..106R} spacetimes. For instance, in \cite{Kiefer:1988ud} the author argues that the classical properties of an intrinsic time variable as the scale factor in FLRW geometry, would emerge in some sense, through the interaction with fermionic degrees of freedom. The semiclassical quantum cosmology was investigated \cite{Louko:1988ay} for the FLRW minisuperspace model with electromagnetic field perturbations around the background. Wavepackets were constructed for the coupling of gravity and a massless, as well as a massive scalar field in the FLRW geometry \cite{Kiefer:1988tr}. Another work with the same underlying geometry and in the presence of scalar field can be found here \cite{Page:1990mh}. For the case of generally, spatially cosmologies in the presence of a scalar field, exact solutions have been found \cite{Christodoulakis:1991jd}. In the case that the Hamilton-Jacobi equation is separable for the cosmological models under consideration, it has been proven \cite{Simeone:1998ni} that these models can be quantized as ordinary gauge systems. The solutions to the Wheeler-DeWitt equation for the Reissner-Nordstr\"{o}m-deSitter black whole where interpretated via the de Broglie-Bohm theory \cite{Kenmoku:1998ax}. The use of conditional symmetries was employed for the quantum description of the general Bianchi Type I, with and without the presence of a cosmological constant \cite{Christodoulakis:2001um}. In the same line of the thought, the canonical quantization of Reissner-Nordstr\"{o}m black hole can be found in \cite{Christodoulakis:2013sya}, where the interpretation of the solutions to the WDW equation is based on the semiclassical corrected geometry via Bohm's analysis. Another work that shares some common ground with what we intent to do in this paper, is \cite{Dimakis:2016mpg}. The authors manage to decouple the reparametrization invariance for some minisuperspace models and hence reduce the systems to the ``true'' degrees of freedom. Furthermore, they provide a generalized definition of probability. For an interpretation of the wavefunctions through the definition of homothetic time we provide this work \cite{Karagiorgos:2018gkn}. A different perspective on the quantum mechanical corrections to the Schwarzschild black hole can be found in \cite{2017EL....11760006B}. The authors study the contributions to the Bekenstein black hole entropy due to quantum corrections. Yet another different approach can be found in \cite{Corda:2019vuk} and \cite{Corda:2020fjz} were a Schr\"{o}dinger type equation was found for the Schwarzschild and Reissner-Nordstr\"{o}m black holes correspondingly. The quantum black holes were treated as a gravitational hydrogen atom and their energy spectrum was found. The authors in \cite{Davidson:2014tda} describe a quantum Schwarzschild black hole by a non-singular wave packet composed of plane wave eigenstates of the momentum Dirac-conjugate to the mass operator. Furthermore, in \cite{Davidson:2012dt}, they provide an interpretation to an emergent spacetime as the quantum mechanical Schwarzschild vacuum which appears to be massless but exhibits non-zero mass uncertainty. These are only few examples of the full list of quantum minisuperspace models.

The purpose of this work is to provide some sort of a Schr\"{o}dinger type equation for the minisuperspace Reissner-Nordstr\"{o}m black hole, without imposing any gauge conditions and without introducing additional functional $\grx^{A}$ for the purpose of choosing ``time'' variables. To do so, we recognize that one of the sources to the various problems it is the existence of constraint equations and hence the singular nature of the Lagrangian density. Our intention is to satisfy the constraint equation without imposing any gauge condition and then construct a reduced Lagrangian density which will be regular for some of the degrees of freedom and furthermore will not have a square root form. Since the gauge is not broken, we expect non-true degrees of freedom to be included in the Lagrangian density. Those will help us to identify the ``time'' parameter and hence to define a Schr\"{o}dinger type equation. For whatever wavefunctions to be found, we intent to use the Bohm analysis in order to present an interpretation based on geometrical tools. Furthermore, a some sort of criterion it is provided with which we adjudicate which, between two spacetimes is ``more'' singular in comparison. Once the Hilbert space is constructed, a probability density can be defined, to which we apply DeWitt's idea on whether singularities are avoided (The probability density should be zero at the configuration points where the classical singularity appears). 
 
The paper is structured as follows: There are four basic sections, the classical description II, the quantum description III, the Bohm Analysis IV and the Discussion V. Related to the section II: the starting point II.1 is the introduction to the line element and the electromagnetic potential for static spherically-symmetric spacetimes, prior to the solution of the equations. In II.2 the well known singular Lagrangian density is provided. The reduced, ``time'' dependent, regular Lagrangian density is presented in II.3. The last subsection II.4 is dedicated to some variable transformations in the minisuperspace geometry and obtaining the solution in arbitrary gauge. When it comes to the quantum description: Some variable redefinition are presented in III.1 and the canonical Hamiltonian density is constructed in III.2. In III.3 the ``time''-covariant Schr\"{o}dinger equation is defined. Two subsections follow, III.3.1, III.3.2, where wavefunctions to the above equation are found for specific initial states. Next is the IV where the Bohm analysis is performed IV.1, IV.2, for the two wavefunctions obtained in the previous section. The singularity criterion is presented in IV.3 for the comparison of the classical and semi-classical trajectories. Finally, the section IV.4 is related to the existence of event horizons for the semi-classical solutions. Lastly, in section V a discussion can be found for the overall results. 
%%%%%%%%%%%%%%%%%%%%%%%%%%%%%%%%%%%%%%%%%%
\section{The classical description}

\subsection{Static, spherically-symmetric spacetime}

The starting point is the Einstein's-Hilbert's-Maxwell's action describing gravity in the presence of free electromagnetic field
\begin{align}
S=\frac{1}{c}\int{\sqrt{-g}dx^{4}\left(\frac{1}{2\grk}R-\frac{1}{4\grm_{0}}F\right)},\label{eq1}
\end{align} 
where $R=R_{\grm\grn}g^{\grm\grn}$ and $F=F_{\grm\grn}F^{\grm\grn}$, with $R_{\grm\grn},g_{\grm\grn},F_{\grm\grn}$ the Ricci, the metric and the Faraday tensor respectively. Furthermore, note that c is the speed of light in vacuum, $\grm_{0}$ the magnetic permeability and $\grk=\frac{8 \grp G}{c^{4}}$, with $G$ the Newton's gravitational constant. The system of equations in tensor component form read
\begin{align}
&G_{\grm\grn}=\grk T_{\grm\grn},\label{eq2}\\
&\nabla_{\grn}F^{\grm\grn}=0,\label{eq3}
\end{align}
where 
\begin{align}
&G_{\grm\grn}=R_{\grm\grn}-\frac{1}{2}R g_{\grm\grn},\label{eq4}\\
&T_{\grm\grn}=\frac{1}{\grm_{0}}\left(F_{\grm\grs}{F_{\grn}}^{\grs}-\frac{1}{4}F g_{\grm\grn}\right),\label{eq5}\\
&F_{\grm\grn}=\nabla_{\grm}A_{\grn}-\nabla_{\grn}A_{\grm},\label{eq6}
\end{align}
and $A_{\grm}=\left(-\frac{\grF}{c},A_{i}\right)$, the electromagnetic four-potential, $(\grm,\grn=0,1,2,3)$,\,$(i=1,2,3)$.\\

The restriction to the desired sub-family of spacetimes, (static, spherically-symmetric) can be realized by the demand that the metric admits the following Lie algebra 
\begin{align}
&\grj_{1}=\sin \grf \partial_{\gru}+\cos \grf \cot \gru \partial_{\grf},\,
\grj_{2}=\cos \grf \partial_{\gru}-\sin \grf \cot \gru \partial_{\grf},\,
\grj_{3}=\partial_{\grf},\,
\grj_{4}=\partial_{t}\label{eq10}.
\end{align}
To this end, the coordinates $(t,r,\gru,\grf)$ have been chosen in such a manner, so that without loss of generality, the line element acquires the form
\begin{align}
ds_{(4)}^{2}=-a(r)^{2}dt^{2}+n(r)^{2}dr^{2}+b(r)^{2}\left(d\gru^{2}+\sin^{2}\gru d\grf^{2}\right).\label{eq11}
\end{align} 
In the light of \eqref{eq11}, all the off-diagonal terms of the Einstein's tensor are equal to zero. When it comes to the electromagnetic potential, the following properties may be used to track down its non-zero components
\begin{enumerate}
\item ``Inheritance'' of the isometries
\begin{align}
\forall \grj_{I}:{\cal{L}}_{\grj_{I}}g_{\grm\grn}=0\Rightarrow {\cal{L}}_{\grj_{I}}G_{\grm\grn}=0\Rightarrow{\cal{L}}_{\grj_{I}}T_{\grm\grn}=0.\label{eq12}
\end{align}
\item Algebraic conditions for this specific metric
\begin{align}
&\forall \grm,\grn=0,1,2,3:\grn>\grm\Rightarrow\,G_{\grm\grn}=0\Rightarrow T_{\grm\grn}=0,\label{eq13}\\
&G_{44}=\sin^{2}\gru G_{33}\Rightarrow T_{44}=\sin^{2}\gru T_{33}.\label{eq14}
\end{align}
\item Electromagnetic gauge freedom
\begin{align}
\tilde{A}_{\grm}=A_{\grm}+\nabla_{\grm}\grL.\label{eq15}
\end{align}
\end{enumerate}
Each one of the above properties introduces no additional ansatz, hence without loss of generality the electromagnetic potential acquires the form
\begin{align}
A_{\grm}=\left(-\frac{\grF(r)}{c},0,0,\grB\cos\gru\right),\label{eq16}
\end{align}
where $\grB$ some arbitrary constant, representing a Dirac monopole. For the purposes of this work, we assume that this constant is equal to zero, hence
\begin{align}
A_{\grm}=\left(-\frac{\grF(r)}{c},0,0,0\right).\label{eq161}
\end{align}

\subsection{Singular, minisuperspace Lagrangian}

The degrees of freedom describing this system are $a(r),n(r),b(r)$ and $\grF(r)$. However, not all of them are ``true''. To unveil the remaining freedom, consider some coordinate transformation of the form $r=f(\tilde{r})$. Due to this, the degrees of freedom transform as
\begin{align}
a(f(\tilde{r}))=\tilde{a}(\tilde{r}),\,n(f(\tilde{r}))\frac{d f(\tilde{r})}{d\tilde{r}}=\tilde{n}(\tilde{r}),\,b(f(\tilde{r}))=\tilde{b}(\tilde{r}),\,\grF(f(\tilde{r}))=\tilde{\grF}(\tilde{r}).\label{eq17}
\end{align}  
The transformation law of a scalar field corresponds to $(a,b,\grF)$ and that of a density to $n$. To this end, the remaining freedom can be used to set, any of the $(a,b,\grF)$ equal to some function of $r$ but not constant, while for $n$, even a constant value is acceptable. The usual ``gauge'' choice, as it is called, is $b(r)=r^{2}$ such that the line element at $(t=constant, r=\gra=constant)$ is equal to that of a sphere with radius $\gra$. 

In this work however, no gauge choice will be made prior to the solution of the equations. Due to this remaining freedom, the system of Einstein's-Maxwell's equations consist of three dynamical equations, meaning three, second order, ordinary, differential equations which can brought into the form
\begin{align}
&\ddot{a}=-\frac{a n^{2}}{2b^{2}}-\frac{\dot{a}\dot{b}}{b}+\frac{a\dot{b}^{2}}{2b^{2}}+\frac{\dot{a}\dot{n}}{n}+\frac{3\grk\dot{\grF}^{2}}{4c^{2}\grm_{0}a} ,\label{eq18}\\
&\ddot{b}=\frac{n^{2}}{2b}-\frac{\dot{b}^{2}}{2b}+\frac{\dot{b}\dot{n}}{n}-\frac{\grk b \dot{\grF}^{2}}{4c^{2}\grm_{0}a^{2}},\label{eq19}\\
&\ddot{\grF}=\frac{\dot{a}\dot{\grF}}{a}-\frac{2\dot{b}\dot{\grF}}{b}+\frac{\dot{n}\dot{\grF}}{n},\label{eq20}
\end{align} 
and a constraint equation 
\begin{align}
-\frac{n^{2}}{b^{2}}+\frac{2\dot{a}\dot{b}}{ab}+\frac{\dot{b}^{2}}{b^{2}}+\frac{\grk \dot{\grF}^{2}}{2c^{2}\grm_{0}a^{2}}=0,\label{eq21}
\end{align}
where no second derivative appears. At the same time, no second derivative of $n$ appears to none of the equations. If someone could solve the dynamical equations, then the constraint equation would impose conditions upon the solutions. 

In trying to construct a Lagrangian density from which we can reproduce  \eqref{eq18}-\eqref{eq21} via the Euler's-Lagrange's variational method, the form of the metric \eqref{eq11} and the potential \eqref{eq161} can be used in the Einstein's-Hilbert's-Maxwell's action. Due to only the $r$-dependence of the degrees of freedom and by dropping out some total derivative terms, the action can be written as follows
\begin{align}
S=\int{Vdx^{3}}\int{L(n,a,\dot{a},b,\dot{b},\grF,\dot{\grF})dr}.\label{eq22}
\end{align}
Since upon integration, the term $\int{Vdx^{3}}$ is just a multiplicative constant (this constant is always infinite due to dt, but will not bother us further), the desired form of the minisuperspace Lagrangian density to reproduce the equations is
\begin{align}
L=\frac{1}{2n}G_{\grm\grn}(q)\dot{q}^{\grm}\dot{q}^{\grn}-n V(q),\label{eq23}
\end{align}
where $q^{\grm}=(a,b,\grF)$,\,$\dot{q}=\frac{d q^{\grm}}{dr}$, $G_{\grm\grn}(q)$ the corresponding minisuperspace metric
\begin{align}
G_{\grm\grn}(q)=\begin{pmatrix}
0 & \frac{2}{\grk}b & 0 \\
\frac{2}{\grk}b & \frac{2}{\grk}a & 0 \\
0 & 0 &\frac{1}{c^{2}\grm_{0}}\frac{b^{2}}{\gra}
\end{pmatrix},\label{eq24}
\end{align}
and the potential 
\begin{align}
V(q)=-\frac{1}{\grk}a.\label{eq25}
\end{align}
The singular nature of this Lagrangian density is related to the absence of $\dot{n}$, hence, the constraint equation is obtained due to the variation with respect to $n$. As we have pointed out in the introduction, some of the arising problems are due to the singular nature of this Lagrangian density and many discussions are focused on the ``best'' way to treat the constraints in the quantization procedure.

\subsection{Reduced, regular, ``time''-dependent Lagrangian density}

What we intend to do in this work, it is to reduce the system of equations into regular, hence to construct a regular Lagrangian density, while at the same time, maintain the covariance under transformations of the form $r=f(\tilde{r})$. The procedure is the following: first solve the constraint equation with respect to the variable $n$. This is easy since the constraint equation is algebraic with respect to the aforementioned variable. To be precise, the expression is simply
\begin{align}
n=\frac{1}{\sqrt{2\grm_{0}}c}\frac{1}{a}\sqrt{4c^{2}\grm_{0}ab\dot{a}\dot{b}+2c^{2}\grm_{0}a^{2}\dot{b}^{2}+\grk b^{2}\dot{\grF}^{2}}.\label{eq26}
\end{align}
As a result, the equation \eqref{eq21} is identically satisfied, while the other three \eqref{eq18}-\eqref{eq20} are reduced. In particular, only two of them are independent,
\begin{align}
&\ddot{a}=-\frac{\dot{a}^{2}}{a}-\frac{2\dot{a}\dot{b}}{b}+\frac{\grk}{2\grm_{0}c^{2}}\frac{\dot{\grF}^{2}}{a}+\dot{a}\frac{\ddot{b}}{\dot{b}},\label{eq27}\\
&\ddot{\grF}=-\frac{2\dot{b}\dot{\grF}}{b}+\dot{\grF}\frac{\ddot{b}}{\dot{b}}.\label{eq28}
\end{align}
For the usual gauge choice $b(r)=r$, the solution to the above equations read
\begin{align}
\grF=-\frac{\grg_{1}}{r}+\grg_{2},\,\gra=\frac{\sqrt{\grg_{1}^{2}\grk+4c^{2}r(\grg_{3}+\grg_{4}r)\grm_{0}}}{\sqrt{2}c\sqrt{\grm_{0}}r},
\end{align}
where $\grg_{1},\grg_{2},\grg_{3},\grg_{4}=$constants. The following choice $\grg_{2}=0$ implies that at $r\rightarrow\infty$ the potential is equal to zero. Furthermore, by the redefinition $\grg_{3}=-r_{s}\grg_{4}$, $\grg_{1}=\frac{2c\sqrt{\grg_{4}\grm_{0}}r_{q}}{\sqrt{\grk}}$ and the coordinate transformation $t=\frac{\tilde{t}}{\sqrt{2\grg_{4}}}$, the well known Reissner-Nordstr\"{o}m solution is obtained
\begin{align}
&g_{\grm\grn}=\begin{pmatrix}
-\left(1-\frac{r_{s}}{r}+\frac{r_{q}^{2}}{r^{2}}\right) & 0 & 0 & 0\\
0 & \frac{1}{\left(1-\frac{r_{s}}{r}+\frac{r_{q}^{2}}{r^{2}}\right)} & 0 & 0\\
0 & 0 & r^{2} & 0\\
0 & 0 & 0 & r^{2}\sin^{2}\gru\\
\end{pmatrix},\label{eqsdsds51}\\
&A_{\grm}=(-\sqrt{2\frac{\grm_{0}}{\grk}}\frac{r_{q}}{r},0,0,0),\label{eqfdghh52}
\end{align}
where the usual definitions hold $r_{q}^{2}=\frac{Q^{2}\grk}{32\grp^{2}}\grm_{0}c^{2}$, and $r_{s}=\frac{2GM}{c^{2}}$ the Schwarzschild radius.

The following remarks are noteworthy

\textbf{Remarks}
\begin{enumerate}
\item The equation \eqref{eq26} does not constitute a gauge choice. There are basically two ways to understand this: the first one is by noticing that only two equations remain \eqref{eq27}-\eqref{eq28} while there are three degrees of freedom $(a,b,\grF)$. The second one is a more of a straightforward procedure, where a transformation of the form $r=f(\tilde{r})$ should be considered. Hence, the gauge freedom still holds. Based on the form that the equations are written, it has been transferred to the function $b$.
\item If it is not a gauge choice then what is it? It is seems that it is just a ``very good'' parametrization of the original line element in the following sense: If someone was lucky or clever enough, he could have chosen from the beginning, before calculate Einstein's-Maxwell's equations, the parametrization of Lapse as it is given from \eqref{eq26}. Then, no constraint equation would appear. Or better, the constraint equation is trivially satisfied once \eqref{eq27}, \eqref{eq28} are solved.
\item The reduced equations could be solved with respect to either one of the pairs $(\ddot{a},\ddot{\grF})$, $(\ddot{a},\ddot{b})$ or $(\ddot{b},\ddot{\grF})$. We have simply chosen the first.
\end{enumerate}
Let us now move to the reduced Lagrangian density. It turns out, that it is constructed from the \eqref{eq23}, after the replacement \eqref{eq26}
\begin{align}
L=\frac{1}{c\grk\sqrt{\grm_{0}}}\sqrt{8c^{2}\grm_{0}ab\dot{a}\dot{b}+4c^{2}\grm_{0}a^{2}\dot{b}^{2}+2\grk b^{2}\dot{\grF}^{2}}.\label{eq29}
\end{align}
It is still a singular Lagrangian density for the dynamical variables $(a,b,\grF)$, since the determinant of the following matrix $A$, is zero
\begin{align}
&A=\frac{\partial L}{\partial \dot{q}^{\grm}\dot{q}^{\grn}},\label{eq30}\\
&Det(A)=0.\label{eq31}
\end{align}
However, since the constraint equation is identically satisfied upon the solutions to the dynamical equations, this intrigue us to think the variable $b$ not as a dynamical degree of freedom, but rather as a ``time'' ($r$)-dependent function. The variation with respect to the degrees of freedom $(a,\grF)$ yields exactly the equations \eqref{eq27}, \eqref{eq28}. Under the previous assumption however, the Lagrangian is now regular but ``time''-dependent. Note that the word time is used only nominally since the variable $r$ is a spatial coordinate. The reason for call it ``time'' will become more clear in the preceding sections.

Still, due to the square root, the canonical quantization procedure would be difficult enough for this sort of Lagrangian density. In order to circumvent this problem, we search for a Lagrangian density which reproduces the correct equations and has the following form
\begin{align}
L=\frac{1}{2}h(b,\dot{b})G_{\grm\grn}(q)\dot{q}^{\grm}\dot{q}^{\grn},\label{eq32}
\end{align}
where $q^{\grm}=(a,\grF)$. One might guess this form as follows: there is no potential part in the equations \eqref{eq27}, \eqref{eq28} and thus no potential is needed in the Lagrangian density. Also, they are linear with respect to the second derivatives for the dynamical degrees of freedom, hence a quadratic dependence on the ``velocities'' would suffice. Furthermore, particular combination of the variables $b,\dot{b}$ holds a role similar to the Lapse function. To this end, the Lagrangian density reads
\begin{align}
L_{(r)}=\frac{1}{2}\frac{b^{2}}{\dot{b}}G_{\grm\grn}\dot{q}^{\grm}\dot{q}^{\grn},\label{eq33}
\end{align} 
where 
\begin{align}
G_{\grm\grn}=\begin{pmatrix}
a^{2} & -\frac{\grk}{2\grm_{0}c^{2}}a\grF\\
-\frac{\grk}{2\grm_{0}c^{2}}a\grF & \frac{\grk^{2}\grF^{2}-8\grm_{0}^{2}c^{4}}{4\grm_{0}^{2}c^{4}}
\end{pmatrix}.\label{eq34}
\end{align}
The particular combination $\frac{b^{2}}{\dot{b}}$ ensures that the Lagrangian density transforms as a density under reparametrizations of $(r)$, meaning that the action $S=\int{L dr}$ is reparametrization invariant and so the Euler-Lagrange equations. For completeness, the fact that the Lagrangian density is regular in the space of $(a,\grF)$ is proven by
\begin{align}
Det\left(\frac{\partial L}{\partial \dot{q}^{\grm}\dot{q}^{\grn}}\right)=-\frac{2a^{2}b^{4}}{\dot{b}^{2}}.\label{eq35}
\end{align}

\subsection{Minisuperspace geometry and variable transformations. The general solution in arbitrary gauge.}

The minisuperspace metric \eqref{eq34} corresponds to flat space with Lorentz signature. This implies that there exist variables $(u,w)$ such that, it acquires the simple diagonal form of a flat Minkowski metric. The transformation needed is
\begin{align}
&a=\frac{\sqrt{\grk u^{2}+8\grm_{0}c^{2}w}}{2c\sqrt{\grm_{0}}},\label{eq36}\\
&\grF=\frac{u}{\sqrt{2}}.\label{eq37}
\end{align}
Furthermore, one may notice that we can define a new function $m(r)$ such that 
\begin{align}
\frac{b^{2}(r)}{\dot{b}(r)}=\frac{1}{m(r)}\Rightarrow b(r)=-\frac{1}{\int{m(r)dr}},\label{eq38}
\end{align}
Due to \eqref{eq38}, the new function $m(r)$ transforms as a Lapse function. In view of these transformations, the Lagrangian density acquires the form
\begin{align}
L=\frac{1}{2m(r)}\tilde{G}_{\grm\grn}\dot{q}^{\grm}\dot{q}^{\grn},\label{eq39}
\end{align} 
where 
\begin{align}
\tilde{G}_{\grm\grn}=\begin{pmatrix}
-1 & 0\\
0 & 1\\
\end{pmatrix},\,q^{\grm}=(u,w).\label{eq40}
\end{align}
The most important outcome of these variable transformations is the fact that the equations \eqref{eq27}, \eqref{eq28} transform into the simple form
\begin{align}
&\ddot{u}=\frac{\dot{m}}{m}\dot{u},\,\ddot{w}=\frac{\dot{m}}{m}\dot{w}.\label{eq42}
\end{align}
What we have actually done, is to take advantage of the knowledge about the geometry of the minisuperspace, and search for variables in which the original metric and the electromagnetic potential could be expressed, in order for the system of Einstein's-Maxwell's equations to acquire the simple form \eqref{eq42}. For completeness, the spacetime metric and the electromagnetic potential in terms of these new variables, have the following non-zero components
\begin{align}
&g_{00}=-\left(\frac{\grk u^{2}}{4\grm_{0}c^{2}}+2w\right),\label{eq43}\\
&g_{11}=\frac{m^{2}\left(\grk u^{2}+8\grm_{0}c^{2}w\right)+\grk\left(\int{mdr}\right)^{2}\dot{u}^{2}-2m\int{mdr}\left(\grk u \dot{u}+4\grm_{0}c^{2}\dot{w}\right)}{\left(\grk u^{2}+8\grm_{0}c^{2}w\right)\left(mdr\right)^{4}},\label{eq44}\\
&g_{22}=\frac{1}{\left(\int{mdr}\right)^{2}}=\frac{1}{\sin^{2}\gru}g_{33},\label{eq45}\\
&A_{0}=-\frac{1}{\sqrt{2}c}u.\label{eq46}
\end{align}
Obviously, the expressions are ``uglier'', but we can tolerate that in light of the simple form of the equations to be solved.
Note that these variable transformations, transform the equations in that form independently of the minisuperspace Lagrangian density that we use to describe our system. The Lagrangian density \eqref{eq33} just helped us to reveal them.

There is another noteworthy fact. The same form of equations has been found in the work \cite{Pailas:2019abb} which describes 3D electromagnetic pp-waves spacetimes. What is the reason behind this occurrence? One reason might be that both spacetimes are Ricci flat, meaning that $R=0$. This holds in the 4D spacetime that we consider here, due to the vanishing of the trace of the electromagnetic energy-momentum tensor $T=g^{\grm\grn}T_{\grm\grn}=0$. This is a property of electromagnetism that holds in any spacetime, when the dimension is equal to four. On the other hand, in 3D, this is a property of pp-wave spacetimes, thus, the condition $T=0$ is imposed. 

To this end, we can obtain the general solution of the charged, static, spherically symmetric, spacetime, in arbitrary gauge, by solving an algebraic equation \eqref{eq21} and an equation of the form
\begin{align}
\ddot{y}=\frac{\dot{m}}{m}\dot{y}.\label{eq47}
\end{align}
This equation can be simply integrated since it can be recast into the form
\begin{align}
&\frac{d}{dr}\ln \dot{y}=\frac{d}{dr}\ln m\Rightarrow\nonumber\\
&\ln \dot{y}=\ln m+\grm_{1}\Rightarrow\nonumber\\
&\dot{y}=\grm_{2}m\Rightarrow\nonumber\\
&y(r)=\grm_{3}+\grm_{2}\int{m(r)dr}.\label{eq48}
\end{align}
Therefore, 
\begin{align}
&u(r)=c_{2}+c_{1}\int{m(r)dr},\,w(r)=c_{4}+c_{3}\int{m(r)dr},\label{eq49}
\end{align}
where $(c_{1},c_{2},c_{3},c_{4})$ are integration constants. We can now turn back to the function $b(r)$, via the relation \eqref{eq38}, since the components of the metric look like more familiar in terms of this arbitrary function. With the following choices of the constants
\begin{align}
c_{2}=0,\,c_{1}=\sqrt{2}\frac{Q\sqrt{2c_{4}}}{4\grp \gre_{0}},\,c_{3}=r_{s} c_{4},\label{eq50}
\end{align}
and the transformation $t=\frac{-\tilde{t}}{\sqrt{2c_{4}}}$, the spacetime metric and the electromagnetic potential acquire the form
\begin{align}
&g_{\grm\grn}=\begin{pmatrix}
-\left(1-\frac{r_{s}}{b(r)}+\frac{r_{q}^{2}}{b(r)^{2}}\right) & 0 & 0 & 0\\
0 & \frac{\dot{b}(r)^{2}}{\left(1-\frac{r_{s}}{b(r)}+\frac{r_{q}^{2}}{b(r)^{2}}\right)} & 0 & 0\\
0 & 0 & b(r)^{2} & 0\\
0 & 0 & 0 & b(r)^{2}\sin^{2}\gru\\
\end{pmatrix},\label{eq51}\\
&A_{\grm}=(-\sqrt{2\frac{\grm_{0}}{\grk}}\frac{r_{q}}{b(r)},0,0,0)\label{eq52}
\end{align}
The constant $c_{2}$ was chosen to be equal to zero because it appeared as an additive constant in the electromagnetic potential, and the demand is that for $b(r)\rightarrow \infty$ the potential must tent to zero $A_{\grm}\rightarrow 0$.

We find that this is one of the simplest ways, to the best of our knowledge, to obtain the Reissner-Nordstr\"{o}m black hole solution.

\section{The quantum description}

\subsection{Variables redefinition}

This section is dedicated to the quantization procedure that we are going to follow in this work. However, before we do so, it is instructive to study the dimensional analysis of the objects defined so far and try to redefine them in a proper way. Since the ``position'' variables that will appear in the Hamiltonian density operators are $(u,w)$, it is necessary to define a new pair $(\grt,\grs)$ which have units of length. Two characteristic length scales that appear in the classical solution are the $(r_{s},r_{q})$. Thus, both can be used to define the new variables. Note that, $w$ is dimensionless, while $u$ share the units of the electromagnetic potential. This can be found form \eqref{eq37}. To this end, the proper redefinitions are
\begin{align}
&u=2c\sqrt{\frac{\grm_{0}}{\grk}}\frac{\grt}{r_{q}},\,w=\frac{1}{2}\frac{\grs}{r_{s}}.\label{eq53}
\end{align}
Alongside, the $m(r)$ will be redefined such that, the Lagrangian describing the dynamics of this system has units of energy. In the usual gauge, $m(r)\sim \frac{1}{[length]^{2}}$, hence we introduce arbitrary dimensional constants as follows
\begin{align}
m=\frac{M_{p}c^{2}}{r_{p}^{2}M},\label{eq54}
\end{align}
where $M_{p}c^{2}\sim [energy]$, $r_{p}\sim[length]$ and $M\sim [energy]$. Hence, the old Lagrangian density, can be written as product of a constant and a new Lagrangian density
\begin{align}
&L=\frac{r_{p}^{2}}{M_{p}c^{2}}\tilde{L},\,\tilde{L}=\frac{M(r)}{2}\left(-\dot{\grt}^{2}+\dot{\grs}^{2}\right).\label{eq55}
\end{align}
For simplicity, the tilde from the Lagrangian density is dropped. The equations of motion do not depend on any of the previously introduced constants, due to the existence of a homothetic vector field for the minisuperspace metric. Furthermore, the only change that appears is a minus sign due to the inversely proportional transformation of m, meaning $m\sim\frac{1}{M}$. To summarize, 
\begin{align}
&L=\frac{M(r)}{2}\left(-\dot{\grt}^{2}+\dot{\grs}^{2}\right),\label{eq56}\\
&\ddot{\grt}=-\frac{\dot{M}}{M}\dot{\grt},\,\ddot{\grs}=-\frac{\dot{M}}{M}\dot{\grs},\label{eq58}
\end{align}
where $M\sim[energy]$, $(\grt,\grs)\sim[length]$, $(\dot{\grt},\dot{\grs})\sim[dimensionless]$ and $L\sim[energy]$. This looks like the Lagrangian density of a free relativistic particle with ``time''-dependent variable mass term. Note however that this is not like the dissipative systems with time-dependent mass since, the $M(r)$ has a different transformation law, due to its geometric origin. The mass in dissipative systems transforms as a scalar, while here, $M(r)$ transforms as a density.

\subsection{Canonical Hamiltonian density and conserved charges}

The procedure to follow is the canonical quantization, hence the canonical Hamiltonian density is defined based on the Legendre transformation. The canonical momenta are defined as usual
\begin{align}
&p_{\grt}=-M\dot{\grt},\,p_{\grs}=M\dot{\grs},\label{eq59}\\
&\dot{\grt}=-\frac{p_{\grt}}{M},\,\dot{\grs}=\frac{p_{\grs}}{M}.\label{eq60}
\end{align}
The canonical Hamiltonian density reads simply
\begin{align}
H_{can}=\frac{1}{2M}\left(-p_{\grt}^{2}+p_{\grs}^{2}\right).\label{eq61}
\end{align}
and the evolution of some quantity $B$ in the phase space, with respect to the parameter $r$ will be given by
\begin{align}
\frac{d\,B}{d r}=\frac{\partial\,B}{\partial r}+\left\{B,H_{can}\right\},\label{eq62}
\end{align}
where $B=B(r,\grt,\grs,p_{\grt},p_{\grs})$. Thus, the phase space equations of motion are
\begin{align}
\dot{\grt}=-\frac{p_{\grt}}{M},\,
\dot{\grs}=\frac{p_{\grs}}{M},\,
\dot{p}_{\grt}=0,\,
\dot{p}_{\grs}=0.\label{eq63}
\end{align}
Due to the explicit $r$ dependence, the canonical Hamiltonian density is not conserved,
\begin{align}
&\frac{d H_{can}}{d r}=\frac{\partial H_{can}}{\partial r}+\left\{H_{can},H_{can}\right\}\Rightarrow\nonumber\\
&\frac{d H_{can}}{d r}=\frac{\partial H_{can}}{\partial r}\Rightarrow\nonumber\\
&\frac{d H_{can}}{d r}=-\frac{1}{2M^{2}}\dot{M}(-p_{\grt}^{2}+p_{\grs}^{2})\Rightarrow\nonumber\\
&\dot{H}_{can}=-\frac{\dot{M}}{M}H_{can}\Rightarrow\nonumber\\
&\frac{d}{dr}\left(M H_{can}\right)=0.\label{eq64}
\end{align}
As we can see, a conserved quantity exists, $H_{con}$,
\begin{align}
&H_{con}=MH_{can}\Rightarrow\nonumber\\
&H_{con}=\frac{1}{2}\left(-p_{\grt}^{2}+p_{\grs}^{2}\right).\label{eq65}
\end{align}
Additionally, due to Noether's theorem, each of the Killing vector fields of the minisuperspace metric is related to a conserved charge, which in our case reads
\begin{align}
Q_{1}=p_{\grt},\,
Q_{2}=p_{\grs},\,
Q_{3}=\grs\,p_{\grt}+\grt\,p_{\grs}.\label{eq66}
\end{align}

\subsection{``Time''-covariant Schr\"{o}dinger equation}

As in the usual quantization of a free particle, the canonical momenta and positions are replaced by operators $(q^{\grm}\rightarrow\hat{q}^{\grm},\,p_{\grm}\rightarrow\hat{p}_{\grm})$ which satisfy the property of \textbf{self-adjointness}
\begin{align}
\left<\hat{p}_{j}\grc\lvert\grf\right>=\left<\grc\lvert\hat{p}_{j}^{*}\grf\right>,\,
\left<\hat{x}^{j}\grc\lvert\grf\right>=\left<\grc\lvert(\hat{x}^{j})^{*}\grf\right>,\label{eq67}
\end{align}
and the \textbf{canonical commutation relations} 
\begin{align}
\left[\hat{x}^{j},\hat{x}^{l}\right]_{C}=0,\,
\left[\hat{x}^{j},\hat{p}_{l}\right]_{C}=i\hbar c \,\grd^{j}_{l},\,
\left[\hat{p}_{j},\hat{p}_{l}\right]_{C}=0,\label{eq68}
\end{align}
where $\left[\cdot,\cdot\right]_{C}$ denotes the commutator, $\grd^{j}_{l}$ the Kronecker's delta, $\left<\grc\lvert\grf\right>=\int{d^{2}x\,\sqrt{|\grm|}\,\grc^{*}\,\grf}$ the inner product and $\sqrt{|\grm|}$ a proper measure. In this case, the chosen one is $\grm=det\left(\tilde{G}_{\grm\grn}\right)$. The position representation suffices for the above properties to be satisfied.
\begin{align}
&\hat{p}_{\grt}=-i \hbar c\partial_{\grt},\,\hat{p}_{\grs}=-i\hbar c\partial_{\grs},\label{eq69}\\
&\hat{\grt}=\grt,\,\hat{\grs}=\grs,\label{eq70}
\end{align}
where the speed of light $c$ in the momentum operator was used for reasons related to unit conventions. In this work, will be interested in wavefunctions that are eigenfunctions of the operators $\hat{H}_{can},\hat{Q}_{1},\hat{Q}_{2}$. The corresponding operators which satisfy the above properties read
\begin{align}
&\hat{H}_{can}=\frac{1}{M}\hat{H}_{con},\label{eq71}\\
&\hat{H}_{con}=-\frac{\hbar^{2}c^{2}}{2}\left(-\partial_{\grt,\grt}+\partial_{\grs,\grs}\right),\label{eq72}\\
&\hat{Q}_{1}=-i\hbar c\partial_{\grt},\label{eq73}\\
&\hat{Q}_{2}=-i\hbar c\partial_{\grs},\label{eq73}
\end{align}
These operators share a common set of eigenfunctions since
\begin{align}
[\hat{H}_{con},\hat{Q}_{1}]=0,\,[\hat{H}_{con},\hat{Q}_{2}]=0,\,[\hat{Q}_{1},\hat{Q}_{2}]=0.\label{eq74}
\end{align}
Note that we do not need to check the commutativity with $\hat{H}_{can}$, it suffices to use $\hat{H}_{con}$, after all, they are related by a multiplicative factor independent of position and momenta.

Let us now proceed to the quantum equation describing this system. As we have pointed out previously, the Hamiltonian density is regular, ``time''-dependent and transforms as a density due to the transformation law of $M(r)$. We have not explicitly chosen a ``time'' variable as can be recognized from the hyperbolic nature of the Hamiltonian density operator. Hence, we will not follow the ``intrinsic'' time that we referred to in the introduction. Furthermore, due to the regularity of the Hamiltonian density, we can define, instead of a Wheeler-DeWitt equation, a Schr\"{o}dinger-like equation and hence witness the evolution of states, in contrast to the ``frozen'' picture related to the quantization of singular systems. What will then be the ``time'' parameter in the Schr\"{o}dinger equation since we have not distinguish any ``time'' variable? The point of view adopted in this work, is that the desired parameter is provided at the point where the $M(r)$ is considered as a function of $r$ and not as a dynamical variable. Thus, $r$ appears as an external parameter in the Hamiltonian density which renders it $r$-dependent. Additionally, the desire to construct a covariant Schr\"{o}dinger equation under ``time'' transformations and thus avoiding the fate of  the ``intrinsic'' time formalism (different ``time'' choices yield different quantum descriptions), narrows down the choice of ``time'' to be the parameter $r$. As a matter of fact, any function of $r$ could serve as a ``time'' parameter, but this freedom is already enclosed in the covariance of the equation. Hence, the desirable parameter is $r$. To this end, the simplest expression is
\begin{align}
i \hbar c \frac{\partial \grC}{\partial r}=\hat{H}_{can}\grC\Leftrightarrow i \hbar c \frac{\partial \grC}{\partial r}=\frac{1}{M}\hat{H}_{con}\grC.\label{eq75}
\end{align}
Note that, due to the transformation law of $M$, originating from its geometric nature, and the transformation of the left hand side, under re-parametrizations of the form $r=f(\tilde{r})$ with arbitrary function $f$, this equation is covariant, in contrast to the usual Schr\"{o}dinger equation which is invariant at most under translations $r\rightarrow\tilde{r}+\gra$, where $\gra=$ some constant. At this point we can also understand why the use of the word ``time'' for the parameter $r$ is only nominal. As we can see from (77) it appears in the place where the time parameter of the usual quantum mechanics would appear.

The solutions of our interest are the common eigenfunctions $\grc(\grt,\grs)$ of the operators $(\hat{H}_{con},\hat{Q}_{1},\hat{Q}_{2})$, hence the system to be solved reads 
\begin{align}
&\hat{H}_{con}\grc=-\frac{E^{2}c^{4}}{2}\grc,\label{eq76}\\
&\hat{Q}_{1}\grc=k_{1}\grc,\label{eq77}\\
&\hat{Q}_{2}\grc=k_{2}\grc,\label{eq78}
\end{align}
where $(E, k_{1}, k_{2})$ some constants. The starting point is to assume that $\grC=T(r)\grc(\grt,\grs)$, hence
\begin{align}
&\eqref{eq75}\Rightarrow i\hbar c \dot{T}(r)\grc(\grt,\grs)=\frac{1}{M(r)}T(r)\hat{H}_{con}\grc(\grt,\grs)\Rightarrow\nonumber\\
&i\hbar c M(r) \frac{\dot{T}(r)}{T(r)}=\frac{\hat{H}_{con}\grc(\grt,\grs)}{\grc(\grt,\grs)}\Rightarrow\\
&i\hbar c M(r) \frac{\dot{T}(r)}{T(r)}=-\frac{E^{2}c^{4}}{2}\Rightarrow
\end{align}
\begin{align}
T(r)=\grn e^{\frac{i}{2\hbar c}E^{2}c^{4}\int{\frac{1}{M(r)}dr}}
,\label{eq79}
\end{align}
where $\grn$ some normalization constant which eventually will be absorbed in the eigenfunctions $\grc$. The eigenfunctions $\grc$ can be determined via the equations \eqref{eq77}, \eqref{eq78}, which yield
\begin{align}
\grc(\grt,\grs)=\frac{1}{2\grp \hbar c} e^{\frac{i}{\hbar c}k_{1}\grt}e^{\frac{i}{\hbar c}k_{2}\grs},\label{eq80}
\end{align}
where the multiplicative factor in front was determined by the normalization of $\grc$ via the Dirac's delta function
\begin{align}
\left<\grc_{k_{1},k_{2}}\lvert\grc_{\tilde{k}_{1},\tilde{k}_{2}}\right>=\grd(k_{1}-\tilde{k}_{1})\grd(k_{2}-\tilde{k}_{2}).\label{eq801}
\end{align}
Use of \eqref{eq80} in \eqref{eq76} and the overall result is 
\begin{align}
&\grC=\frac{1}{2\grp \hbar c} e^{\frac{i}{2\hbar c}E^{2}c^{4}\int{\frac{1}{M(r)}dr}}e^{\frac{i}{\hbar c}k_{1}\grs}e^{\frac{i}{\hbar c}k_{2}\grt},\label{eq81}\\
&E^{2}c^{4}=k_{1}^{2}-k_{2}^{2}.\label{eq82}
\end{align}
As it was expected, the states of this system have well determined momenta $(k_{1},k_{2})$ due to the identifications $\hat{Q}_{1}\leftrightarrow \hat{p}_{\grt}$, $\hat{Q}_{2}\leftrightarrow \hat{p}_{\grs}$. Furthermore, the state is re-parametrization invariant since it came as a solution of an equation which shared this property. 

One of the most important aspects is that we can see the evolution of states. Two cases will be considered, the one being the limit of the other. Given an initial state $\grC_{0}(\grt,\grs)$, recall that the evolution is given by
\begin{align}
\grC_{evo}(r,\grt,\grs)=\int_{-\infty}^{\infty}\int_{-\infty}^{+\infty}{c(k_{1},k_{2})\grC(r,\grt,\grs)} dk_{1}dk_{2},\label{eq83}
\end{align}
where $\grC(r,\grt,\grs)$ the eigenstate \eqref{eq81} and $c(k_{1},k_{2})$ is obtained via
\begin{align}
&c(k_{1},k_{2})=\int_{-\infty}^{\infty}\int_{-\infty}^{+\infty}{\grc^{*}(\grt,\grs)\grC_{0}(\grt,\grs)d\grt d\grs},\label{eq84}
\end{align}
with $\grc^{*}(\grt,\grs)$ the complex conjugate of \eqref{eq80}. 

\subsubsection{Delta-function initial state}

For this subsection the initial state to be considered reads
\begin{align}
\grC_{0}(\grt,\grs)=\grd(\grt-\grt_{0})\grd(\grs-\grs_{0}).\label{eq85}
\end{align}
This implies that the ``position'' has a definite value. This initial state leads to 
\begin{align}
&c(k_{1},k_{2})=\frac{1}{2\grp \hbar c}e^{-\frac{i}{\hbar c}k_{1}\grt_{0}}e^{-\frac{i}{\hbar c}k_{2}\grs_{0}},\label{eq86}\\
&\grC_{\grd}(r,\grt,\grs)=\grV(r,\grt,\grs)e^{\frac{i}{\hbar c}S(r,\grt,\grs)},\label{eq861}\\
&\grV(r,\grt,\grs)=\frac{1}{2\grp \hbar c B(r)},\label{eq862}\\
&S(r,\grt,\grs)=\frac{1}{2 B(r)}\left[-(\grt-\grt_{0})^{2}+(\grs-\grs_{0})^{2}\right],\label{eq87}
\end{align}
where for simplicity the abbreviation $B(r)=\int{\frac{1}{M(r)}dr}$ was defined and will be used from now on wherever it is needed. 
Since there is a well defined Hilbert space, the ``momentum'' probability and the probability density can be defined as well
\begin{align}
&\grP(k_{1},k_{2})= c^{*}(k_{1},k_{2})c(k_{1},k_{2})=\frac{1}{(2\grp \hbar c)^{2}},\label{eq88}\\
&P(r,\grt,\grs) =\grC_{\grd}^{*}(r,\grt,\grs)\grC_{\grd}(r,\grt,\grs)=\frac{1}{(2\grp \hbar c B(r))^{2}}.\label{eq89}
\end{align}
As it was expected, since the ``position'' has a definite value, all the ``momenta'' share the same probability \eqref{eq88}. The probability density is re-parametrization invariant and hence the probability itself will be so.

\subsubsection{Gaussian initial state}

The initial state reads
\begin{align}
\grC_{0}(\grt,\grs)=\sqrt{\frac{2}{\grp \grl^{2}}}e^{-\frac{(\grt-\grt_{0})^{2}}{\grl^{2}}}e^{-\frac{(\grs-\grs_{0})^{2}}{\grl^{2}}}.\label{eq90}
\end{align}
The previous initial state can being seen as a limit of this one at $\grl \rightarrow 0$. Following the same procedure as before we get 
\begin{align}
&c(k_{1},k_{2})=\frac{1}{2\grp \hbar c}\sqrt{2\grp \grl^{2}}e^{-\frac{\grl^{2}}{4\hbar^{2}c^{2}}k_{1}^{2}-\frac{i}{\hbar c}\grt_{0}k_{1}}e^{-\frac{\grl^{2}}{4\hbar^{2}c^{2}}k_{2}^{2}-\frac{i}{\hbar c}\grs_{0}k_{2}},\label{eq91}\\
&\grC_{G}(r,\grt,\grs)=\grV(r,\grt,\grs)e^{\frac{i}{\hbar c}S(r,\grt,\grs)},\label{eq92}\\
&\grV(r,\grt,\grs)=\frac{\grl^{2}}{2\grp \hbar c}\sqrt{2\grp \grl^{2}}\frac{1}{\sqrt{1+\left(\frac{2\hbar c B(r)}{\grl^{2}}\right)^{2}}}e^{-\left[(\grt-\grt_{0})^{2}+(\grs-\grs_{0})^{2}\right]/\grl^{2}\left[1+\left(\frac{2\hbar c B(r)}{\grl^{2}}\right)^{2}\right]},\label{eq93}\\
&S(r,\grt,\grs)=\frac{2 \hbar^{2}c^{2}B(r)}{\grl^{4}\left[1+\left(\frac{2\hbar c B(r)}{\grl^{2}}\right)^{2}\right]}\left[-(\grt-\grt_{0})^{2}+(\grs-\grs_{0})^{2}\right].\label{eq94}
\end{align}
The corresponding probability densities read
\begin{align}
&\grP(k_{1},k_{2})=\frac{2\grp \grl^{2}}{(2\grp\hbar c)^{2}}e^{-\frac{\grl^{2}}{2\hbar^{2}c^{2}}k_{1}^{2}}e^{-\frac{\grl^{2}}{2\hbar^{2}c^{2}}k_{2}^{2}},\label{eq95}\\
&P(r,\grt,\grs)=\grl^{4}\frac{2\grp \grl^{2}}{(2\grp\hbar c)^{2}}\frac{1}{1+\left(\frac{2\hbar c B(r)}{\grl^{2}}\right)^{2}}e^{-2\left[(\grt-\grt_{0})^{2}+(\grs-\grs_{0})^{2}\right]/\grl^{2}\left[1+\left(\frac{2\hbar c B(r)}{\grl^{2}}\right)^{2}\right]}.\label{eq96}
\end{align}
In contrast to the previous case, the ``momenta'' are not equally probable.

\section{Bohm Analysis}

Even though we have a well defined inner product and the solutions at hand, so we could calculate expectation values and so fourth, we find it more insightful to interpret the wavefunctions by their impact on the geometry. That is to say, what are the quantum corrections to the classical geometry.
 
On of the ways to do such a thing is based on Bohm's analysis \cite{PhysRev.85.166,PhysRev.85.180,Bohm1984}. A de Broglie-Bohm interpretation for the full quantum-gravitational system can be found in \cite{Shtanov:1995ie}. For a recent review check \cite{Pinto-Neto:2018zvn}. Let us briefly recall the main idea properly adjusted in our example: Suppose some Hamiltonian density, and a Schr\"{o}dinger equation of the form
\begin{align}
&\hat{H}_{con}=-\frac{\hbar^{2}c^{2}}{2}\nabla_{\grm}\nabla^{\grn}+V,\label{eq97}\\
&i \hbar c \frac{\partial \grC}{\partial r}=\frac{1}{M(r)}\hat{H}_{con}\grC.\label{eq98}
\end{align}
Furthermore, assume that the states can be cast into the form
\begin{align}
\grC(r,q)=\grV(r,q) e^{\frac{i}{\hbar c}S(r,q)},\label{eq99}
\end{align}
where $\grV(r,q),S(r,q)$ real functions respectively called amplitude and phase of the state. By use of \eqref{eq99} in \eqref{eq98} and equate to zero the real and imaginary parts of the resulting expression, two equations come up
\begin{align}
&\frac{\partial \grV^{2}}{\partial r}+\nabla_{\grm}\left[\grV^{2}\frac{\nabla^{\grm}S}{m(r)}\right]=0,\label{eq100}\\
&\frac{\partial S}{\partial r}+\frac{1}{2M(r)}\nabla_{\grm}S\nabla^{\grm}S+V-Q_{p}=0,\label{eq101}
\end{align}
where 
\begin{align}
Q_{p}=\frac{\hbar^{2}c^{2}}{2M(r)}\frac{\nabla_{\grm}\nabla^{\grm}\grV}{\grV},\label{eq102}
\end{align}
the so called ``Quantum potential''. The equation \eqref{eq100} has the form of a continuity equation with $\grV^{2}$ the ``density'' and $\frac{\nabla^{\grm}S}{M(r)}$ the ``velocity'' of the ``fluid''. This is basically the local expression for the probability conservation. Additionally, the \eqref{eq101} is a modified version of the Hamilton-Jacobi equation, by the Quantum potential term. Thus, in the absence of it, the classical Hamilton-Jacobi equation is retrieved, hence the justification of the term ``Quantum potential''. 

Following Bohm, the connection to the classical regime is to identify the ``fluid momenta'' $(p^{(Q)}_{\grm}=\nabla_{\grm}S)$ to the classical one, given by the Lagrangian density describing the system $(p_{\grm}=\frac{\partial L}{\partial \dot{q}^{\grm}}$), that is
\begin{align}
&p^{(Q)}_{\grm}=p_{\grm}\Leftrightarrow\nonumber\\
&\nabla_{\grm}S=\frac{\partial L}{\partial \dot{q}^{\grm}}.\label{eq103}
\end{align}
The equation \eqref{eq103} is a system of first order differential equations for the ``positions'' $q^{\grm}$, and will provide us with what is called the ``semi-classical trajectories''. 

For later purposes, the easiest way to compare the semi-classical trajectories to the classical ones, is to make the usual gauge choice 
\begin{align}
M(r)=-\frac{M_{p}c^{2}}{r_{p}^{2}}r^{2},\label{eq104}
\end{align}
which implies that at $t=constant, r=constant$ the line element acquires the form
\begin{align}
ds_{(2)}^{2}=r^{2}(d\gru^{2}+\sin^{2}\gru d\grf^{2}).\label{eq105}
\end{align}

\subsection{Delta-function initial state}

In this case, the quantum potential is calculated to be equal to zero. This was expected since as we can see from \eqref{eq862} the state's amplitude is independent of the coordinates $(\grt,\grs)$. The continuity equation is satisfied, with a ``current probability density''
\begin{align}
J^{\grm}=-\frac{M_{p}^{2}c^{2}}{4\grp^{2}\hbar^{2}r_{p}^{4}}\left(\grt-\grt_{0},\grs-\grs_{0}\right).\label{eq106}
\end{align}
As it is expected, the equations \eqref{eq103} are of the form
\begin{align}
&\dot{\grt}=-\frac{1}{r}\left(\grt-\grt_{0}\right),\,\dot{\grs}=-\frac{1}{r}\left(\grs-\grs_{0}\right),\label{eq108}
\end{align}
which yield the classical solutions
\begin{align}
&\grt=\grt_{0}+\frac{\grn_{1}}{r},\,\grs=\grs_{0}+\frac{\grn_{1}}{r}.\label{eq110}
\end{align}
As we can see, the constants $(\grt_{0},\grs_{0})$ that were introduced in the initial quantum state, appear as integration constants for the semi-classical trajectory. The most important outcome is that, for the above quantum description, when the initial state is localized, or the wavefunction has infinite ``uncertainty'' in the ``momenta'', the semi-classical solution coincide with the classical one.

\subsection{Gaussian initial state}

The introduction of a Gaussian initial state is expected to give us one of the possible ways to deviate from the classical trajectories, since it approaches the above Delta state only at the limit $\grl\rightarrow0$, as we have already pointed out. For the solutions to be unburdened from unnecessary, arbitrary constants, that were introduced for a matter of unit conventions, we introduce the parameter $l_{p}$ which will be defined instead of $\grl$, via the relation
\begin{align}
&\grl=\sqrt{\frac{2\hbar}{c M_{p}}}r_{p}\frac{1}{\sqrt{l_{p}}},\label{eq111}\\
&\text{or in terms of Planck's length},\nonumber\\
&\grl=\sqrt{2}\frac{r_{p}^{3/2}}{l_{p}^{1/2}}.\label{eq112}
\end{align} 
Hence, the Delta initial state is expected to be retrieved at the limit $l_{p}\rightarrow\infty$ and so the classical trajectories. Note also that $l_{p}$ has units of length.

Now, the quantum potential is not zero in this case
\begin{align}
Q_{p}=\frac{M_{p}c^{2}l_{p}^{2}}{2r_{p}^{2}(l_{p}^{2}+r^{2})^{2}}\left(\grt+\grs-\grt_{0}-\grs_{0}\right)\left(\grt-\grs-\grt_{0}+\grs_{0}\right),\label{eq113}
\end{align}
and the continuity equation is again satisfied with the probability current reading
\begin{align}
J^{\grm}=-\frac{4\hbar r_{p}^{6}r}{\grp c^{5}M_{p}^{3}l_{p}\left(l_{p}^{2}+r^{2}\right)^{2}}Exp\left[-\frac{M_{p}c l_{p}r^{2}\left[(\grt-\grt_{0})^{2}+(\grs-\grs_{0})^{2}\right]}{\hbar r_{p}^{2}\left(l_{p}^{2}+r^{2}\right)}\right]\left(\grt-\grt_{0},\grs-\grs_{0}\right).\label{eq114}
\end{align}
The equations \eqref{eq103} are of the form
\begin{align}
&\dot{\grt}=-\frac{l_{p}^{2}}{r\left(l_{p}^{2}+r^{2}\right)}\left(\grt-\grt_{0}\right),\label{eq115}\\
&\dot{\grs}=-\frac{l_{p}^{2}}{r\left(l_{p}^{2}+r^{2}\right)}\left(\grs-\grs_{0}\right),\label{eq116}
\end{align} 
which are not the same as in the classical case. The semi-classical solutions read
\begin{align}
&\grt=\grt_{0}+\sqrt{1+\left(\frac{r}{l_{p}}\right)^{2}}\frac{\grn_{1}}{r},\label{eq117}\\
&\grs=\grs_{0}+\sqrt{1+\left(\frac{r}{l_{p}}\right)^{2}}\frac{\grn_{2}}{r},\label{eq118}
\end{align} 
where the integration constants $(\grn_{1},\grn_{2})$ are redefined in order to coincide with the corresponding integration constants of the classical solution. Hence, the radii $r_{q},r_{s}$ can be re-introduced. Furthermore, the correspondence between the integration constants of the classical and semi-classical trajectories can be found from the classical limit $l_{p}\rightarrow \infty $, to be $c_{2}\leftrightarrow \grt_{0},c_{1}\leftrightarrow\grn_{1},c_{4}\leftrightarrow \grs_{0},c_{3}\leftrightarrow \grn_{2}$. That being taken into account, the semi-classical metric and electromagnetic potential have the form
\begin{align}
&g_{\grm\grn}=\begin{pmatrix}
-g_{00}(r) & 0 & 0 & 0\\
0 & \frac{1}{g_{00}(r)}\left[1-\frac{\left(\frac{r_{q}}{l_{p}}\right)^{2}}{1+\left(\frac{r}{l_{p}}\right)^{2}}\right]-\frac{1-g_{00}(r)}{g_{00}(r)}\frac{\left(\frac{r}{l_{p}}\right)^{2}}{1+\left(\frac{r}{l_{p}}\right)^{2}} & 0 & 0\\
0 & 0 & r^{2} & 0\\
0 & 0 & 0 & r^{2}\sin^{2}\gru\\
\end{pmatrix},\label{eq119}\\
&A_{\grm}=(-\sqrt{2\frac{\grm_{0}}{\grk}}\sqrt{1+\left(\frac{r}{l_{p}}\right)^{2}}\frac{r_{q}}{r},0,0,0),\label{eq120}
\end{align}
where the function $g_{00}(r)$ reads
\begin{align}
&g_{00}(r)=1-\sqrt{1+\left(\frac{r}{l_{p}}\right)^{2}}\frac{r_{s}}{r}+\left(\frac{r_{q}}{r}\right)^{2}+\left(\frac{r_{q}}{l_{p}}\right)^{2}\label{eq121}.
\end{align}
In addition to the different form of the metric and electromagnetic potential, the \eqref{eq119}, \eqref{eq120} do not satisfy the free Einstein's-Maxwell's equations. An additional energy-momentum and a four-current tensor have to be taken into account,
\begin{align}
&G_{\grm\grn}=\grk \left(T^{(em)}_{\grm\grn}+T^{(Q)}_{\grm\grn}\right),\label{eq122}\\
&\nabla_{\grn}F^{\grm\grn}=\grm_{0}J_{(Q)}^{\grm}.\label{eq123}
\end{align} 
Both $T^{(Q)}_{\grm\grn}, J^{\grm}_{(Q)}$ are consider quantum mechanical in origin. For completeness, their non-zero components are provided
\begin{align}
&T^{(Q)}_{00}=\frac{1}{\grk}\frac{g_{00}(r)^{2}}{l_{p}^{2}}\frac{2\left(\frac{r_{q}}{l_{p}}\right)^{2}-\sqrt{1+\left(\frac{r}{l_{p}}\right)^{2}}\left(\frac{r_{s}}{r}\right)}{\left[1+\left(\frac{r}{l_{p}}\right)^{2}g_{00}(r)-\left(\frac{r_{q}}{l_{p}}\right)^{2}\right]^{2}},\label{eq124}\\
&T^{(Q)}_{22}=\frac{1}{4\grk}\frac{4\left(\frac{r_{q}}{l_{p}}\right)^{2}-\left(\frac{r_{s}}{l_{p}}\right)^{2}}{\left[1+\left(\frac{r}{l_{p}}\right)^{2}g_{00}(r)-\left(\frac{r_{q}}{l_{p}}\right)^{2}\right]^{2}},\label{eq125}\\
&T^{(Q)}_{33}=T_{22}\sin^{2}\gru,\label{eq126}\\
&J^{0}_{(Q)}=-\sqrt{\frac{2}{\grk\grm_{0}}}\frac{r_{q}}{l_{p}^{2}r\sqrt{1+\left(\frac{r}{l_{p}}\right)^{2}}}\frac{1+\left(\frac{r}{l_{p}}\right)^{2}g_{00}(r)-\frac{1}{2}\sqrt{1+\left(\frac{r}{l_{p}}\right)^{2}}\frac{r_{s}}{r}}{\left[1+\left(\frac{r}{l_{p}}\right)^{2}g_{00}(r)-\left(\frac{r_{q}}{l_{p}}\right)^{2}\right]^{2}}.\label{eq127}
\end{align}
By evaluating carefully the limit of the above expressions at $l_{p}\rightarrow\infty$ we notice that all of them tend to zero, as expected. Thus, the realization of the quantization effect is localized in the alteration of the forms of the original fields and appearance of sources, in both sets of equations.

\subsection{Singularity}

Let us now discuss about the singularities. For the classical solution, the Ricci scalar is equal to zero, due to the special behavior of the electromagnetic field in four dimensions, that is, the trace of the energy momentum tensor is equal to zero. This is a generic feature and does not depend on whether or not the solution is classical or semi-classical. As we have seen, for the semi-classical solution, an additional energy momentum tensor has to be taken into account. This leads to a non-zero Ricci scalar which vanishes at the limit $l_{p}\rightarrow\infty$, sharing the same property of the $T^{(Q)}_{\grm\grn}$. The behavior of this scalar for small values of $r$ (we present only the divergent terms) is
\begin{align}
R\simeq -\frac{r_{s}r_{q}^{2}}{l_{p}^{2}}\frac{1}{r^{3}}+\frac{4r_{q}^{2}\left(r_{q}^{2}-r_{s}^{2}\right)+l_{p}^{2}\left(-4r_{q}^{2}+3r_{s}^{2}\right)}{2l_{p}^{4}}\frac{1}{r^{2}}+\frac{rs\left[-2l_{p}^{4}+12r_{q}^{4}-6r_{q}^{2}r_{s}^{2}+l_{p}^{2}\left(-11r_{q}^{2}+6r_{s}^{2}\right)\right]}{2l_{p}^{6}}\frac{1}{r}.\label{eq128}
\end{align}
Thus, we can immediately conclude that this spacetime is not singularity free. Since in the classical case $R=0$, the Ricci scalar is not a good measure of whether this spacetime is ``more'' or ``less'' singular than the classical Reissner-Nordström. Let us explain what we mean by ``more'' or ``less'' singular: 
Consider the cases of Schwarzschild and Reissner-Nordström black hole. Since in both cases $R=0$, we are going to use the Kretschmann scalar $K=R^{\grm\grn\grs\grr}R_{\grm\grn\grs\grr}$ for our purpose, which reads respectively
\begin{align}
&K_{(S)}=\frac{12r_{s}^{2}}{r^{6}},\label{eq130}\\
&K_{(RN)}=\frac{56r_{q}^{4}}{r^{8}}-\frac{48 r_{q}^{2}r_{s}}{r^{7}}+\frac{12r_{s}^{2}}{r^{6}}\label{eq129}.
\end{align}
In order to work with dimensionless objects, we multiply both expressions with $r_{s}^{4}$ and provide $(r_{q},r)$ in terms of $r_{s}$ as well, specifically $r_{q}=\gra \frac{rs}{2}$, $r=x r_{s}$, with ($0\leq\gra<1$ so that no naked singularity appears.) 
\begin{align}
&X_{(RN)} =K_{(RN)}r_{s}^{4}= \frac{7}{2}\frac{\gra^{4}}{x^{8}}-12\frac{\gra^{2}}{x^{7}}+\frac{12}{x^{6}},\label{eq131}\\
&X_{(S)}= K_{(S)}r_{s}^{4}=\frac{12}{x^{6}}.\label{eq132}
\end{align}
Now the difference between the two is defined
\begin{align}
X= X_{(RN)}-X_{(S)}= \frac{7}{2}\frac{\gra^{4}}{x^{8}}-12\frac{\gra^{2}}{x^{7}},\label{eq133}
\end{align} 
and we provide the contour plot in the space of the two parameters $(\gra,x)$. 

\begin{figure}
\centering    
\includegraphics[width=9cm, height=8cm]{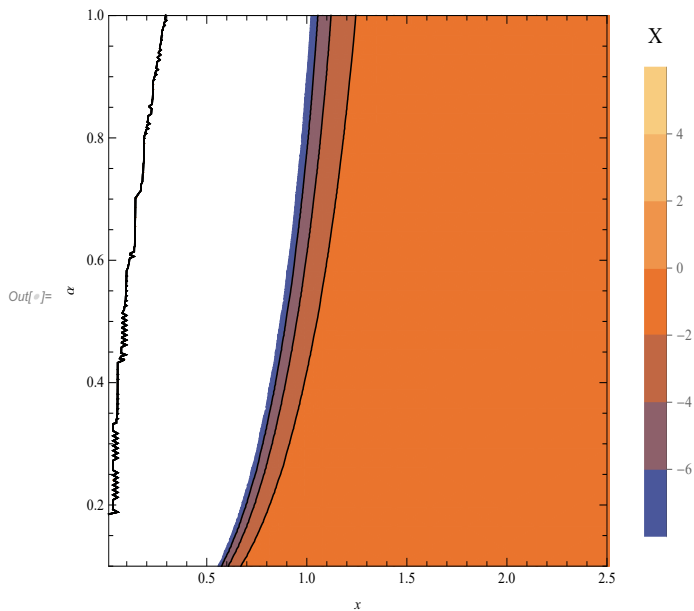}
\caption{This is the contour plot for the function $X$ in the space of parameters $(x,\gra)$.}
\label{Fig. 1} 
\end{figure}
 
As it can be inferred from figure \ref{Fig. 1}, the function $X$ acquires negative values in the accepted parameter space. This means that $|X_{(S)}|>|X_{(RN)}|$, hence we can ``move'' towards smaller values of x (or r) in the case of Reissner-Nordström rather than that of Schwarzschild, before the curvature ``blows up'' to infinity. In some sence, the addition of electromagnetic field (a point charge), leads to a more localized singularity around the value $r=0$ and hence, by this criteria, a ``less'' singular spacetime. 

\begin{figure}
\centering    
\includegraphics[width=14cm, height=6cm]{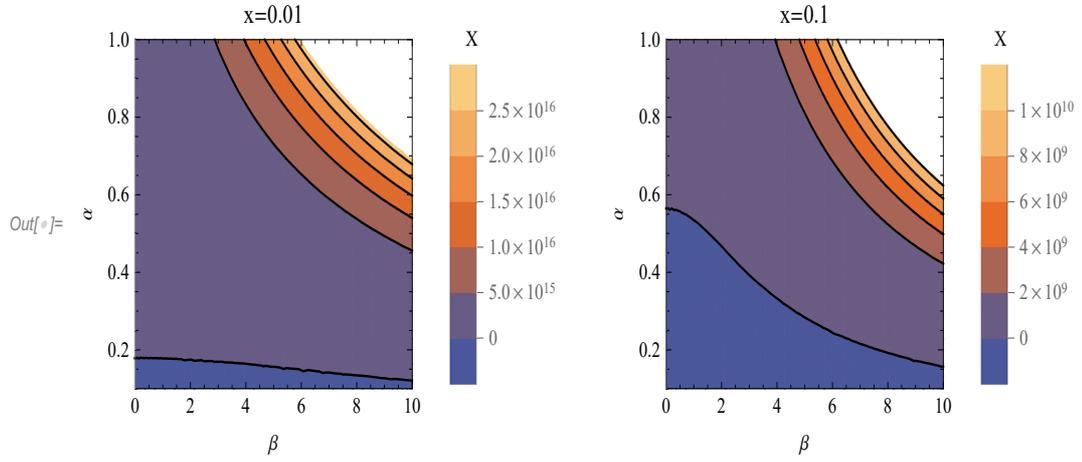}
\caption{These are the contour plots for the function $X$ in the space of parameters $(\gra,\grb)$ for two values of $x$, $(x=0.01),\,(x=0.1)$.}
\label{Fig. 2} 
\end{figure}

\begin{figure}
\centering    
\includegraphics[width=14cm, height=6cm]{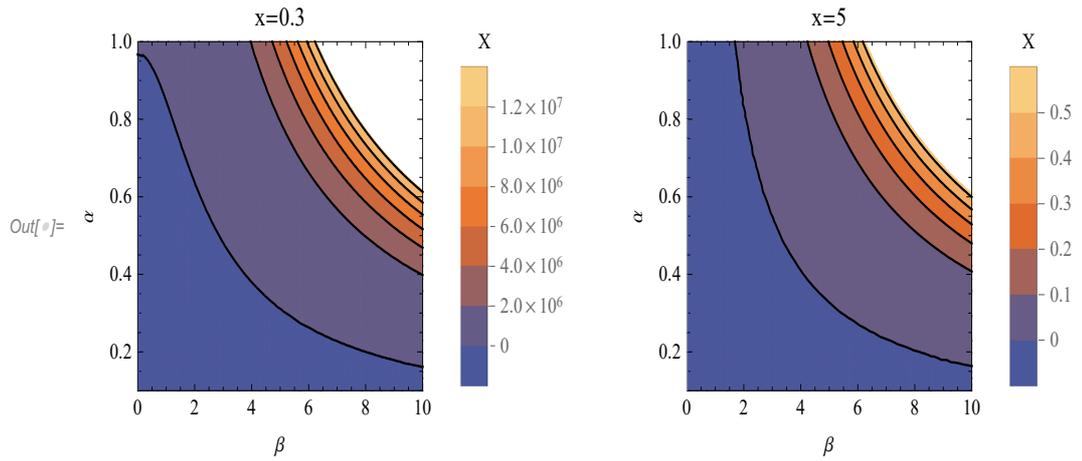}
\caption{These are the contour plots for the function $X$ in the space of parameters $(\gra,\grb)$ for two values of $x$, $(x=0.3),\,(x=5)$.}
\label{Fig. 3} 
\end{figure}

To this end, we are going to compare the classical Reissner-Nordström and the quantum corrected one (QRN) based on the above described method. To proceed, the constant $l_{p}$ is redefined in terms of $r_{s}$ as follows $l_{p}=\frac{r_{s}}{\grb}$, where for $\grb\rightarrow {0}$ the classical Reissner-Nordström is retrieved. The quantum RN has many divergent terms, but we keep only the higher ones, which coincide in powers of $r$ with those of the classical solution. For completeness,
\begin{align}
&X_{(QRN)}=X_{(RN)}+\frac{23}{4}\frac{\left(\gra^{2}\grb\right)^{2}}{x^{7}}+\frac{\left(\gra\grb\right)^{2}}{x^{6}}\left\{-\frac{37}{2}+\gra^{2}\left[\frac{1}{2}+\frac{111}{16}\grb^{2}-\frac{9}{8}\left(\gra\grb\right)^{2}\right]\right\},\label{eq134}\\
&X=\frac{23}{4}\frac{\left(\gra^{2}\grb\right)^{2}}{x^{7}}+\frac{\left(\gra\grb\right)^{2}}{x^{6}}\left\{-\frac{37}{2}+\gra^{2}\left[\frac{1}{2}+\frac{111}{16}\grb^{2}-\frac{9}{8}\left(\gra\grb\right)^{2}\right]\right\}.\label{eq134}
\end{align}
Now the space of parameters is three-dimensional, thus we could give a density plot. However, in trying we found out that it is not very presentable. Thus, we decide to construct some contour plots in the space of variables $(\gra,\grb)$ for some specific values of $x$. The results are presented in the figures \ref{Fig. 2}, \ref{Fig. 3}. 

Some remarks are at hand: The first thing that we observe is that there are values of the parameters that imply $|X_{(QRN)}|>|X_{(RN)}|$, meaning that the quantum corrected spacetime is ``more'' singular than the classical one based on the above criterion. The second thing is that in all four cases, there exists the deep blue region which corresponds to $|X_{(QRN)}|<|X_{(RN)}|$ and hence ``less'' singular spacetime. Finally, as we move closer to $(x=0)$, the desired region becomes smaller. So, we can always achieve ``less'' singular spacetimes for some values of $\grb$ but the characteristics of those black holes (encoded in $\gra$) are bounded to specific values. If we now focus on a specific distance $x$, then we observe that as $\grb$ grows, meaning stronger quantum effects, the region of accepted values of $\gra$ is closer to $\gra=0$, which corresponds to a Schwarzschild black hole.

\subsection{Event horizons}

This solution exhibits event horizons as the classical Reissner-Nordström solution. From \eqref{eq119} the solution $g_{00}(r)=0$ leads to the divergence of the second diagonal element of the metric tensor. There are four solutions to the above equation, but only two of them correspond to a positive sign for $r$. The result is
\begin{align}
&r^{(Q)}_{+}=r_{+}\sqrt{\frac{l_{p}^{2}}{l_{p}^{2}-r_{+}^{2}}},\, r^{(Q)}_{-}=r_{-}\sqrt{\frac{l_{p}^{2}}{l_{p}^{2}-r_{-}^{2}}},\label{eq129}
\end{align}
where $r_{\pm}$ the usual Reissner-Nordström inner and outer horizons
\begin{align}
&r_{+}=\frac{1}{2}\left(r_{s}+\sqrt{-4r_{q}^{2}+r_{s}^{2}}\right),\,
r_{-}=\frac{1}{2}\left(r_{s}-\sqrt{-4r_{q}^{2}+r_{s}^{2}}\right).\label{eq131}
\end{align}
There are some worth mentioning remarks. 
\begin{enumerate}
\item The existence of both inner and outer horizons is possible only if the following relations are true
\begin{align}
l_{p}>r_{+}.\label{eq132}
\end{align}
\item At the case $l_{p}=r_{+}$ there is only one horizon which we are going to call $r_{(+-)}$
\begin{align}
r^{(Q)}_{(+-)}=\frac{r_{+}r_{-}}{\sqrt{r_{+}^{2}-r_{-}^{2}}}\Leftrightarrow r^{(Q)}_{(+-)}=\frac{r_{q}^{2}}{\sqrt{r_{s}}\left(-4r_{q}^{2}+r_{s}^{2}\right)^{1/4}}.\label{eq133}
\end{align}
\item At the case $l_{p}=r_{-}$ there is no horizon at all. Better expressed is that, the value of the coordinate $r$ at which $g_{00}(r)=0$, is imaginary. For this particular value $l_{p}=r_{-}$ the parameter $\grb$ becomes a function of $\gra$, $\grb=\frac{2}{1-\sqrt{1-\gra^{2}}}$. In the absence of horizon, we recognize the existence of a naked point singularity, since as can be inferred from figure \ref{Fig. 4}, there are no values $(\gra,x)$ for which $|X_{(QRN)}|<|X_{(RN)}|$.
\begin{figure}
\centering    
\includegraphics[width=9cm, height=8cm]{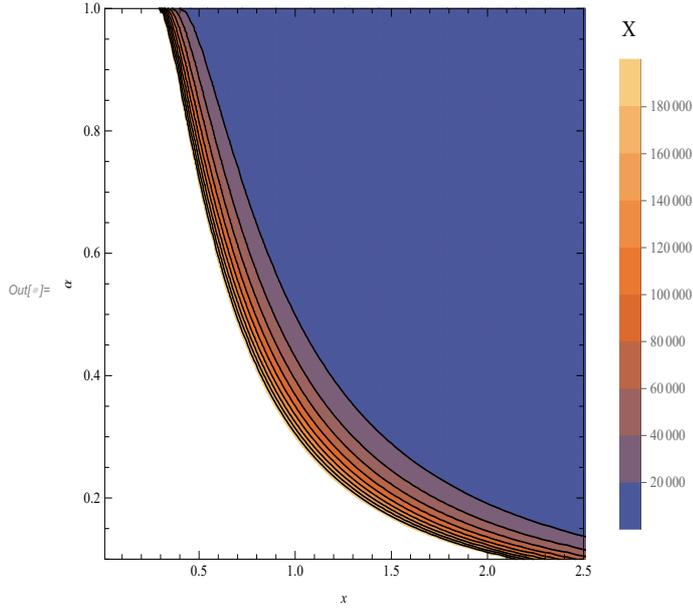}
\caption{This is the contour plot for the function $X$ in the space of parameters $(x,\gra)$.}
\label{Fig. 4} 
\end{figure}
\item From the above two remarks we conclude that, as the quantum effects become stronger, meaning $l_{p}\rightarrow 0$, the horizons are vanishing. For instance, a black hole with $l_{p}>r_{+}$, would consist of a region $r\leq r^{(Q)}_{+}$ that no information could escape. In the case $l_{p}=r_{+}$, this region is defined by $r\leq r^{(Q)}_{+-}$. Since $r^{(Q)}_{+-}<r^{(Q)}_{+}$, information could escape from regions that could not before. 
\item There is also the extremal horizon for which $r_{s}=2r_{q}$ and implies a degeneracy in this case as well
\begin{align}
r^{(Q)}_{+}=r^{(Q)}_{-}.\label{eq134}
\end{align}
This extremal case could not exist for the value $l_{p}=r_{+}$.
\item In the accepted domain of $l_{p}$ in order for $r^{(Q)}_{+},r^{(Q)}_{-}$ to exist, we find that the following inequalities must hold
\begin{align}
&r^{(Q)}_{+}>r_{+},\label{eq135}\\
&r^{(Q)}_{-}>r_{-}.\label{eq136}
\end{align}
Furthermore, an additional holds for $l_{p}>\frac{r_{+}r_{-}}{\sqrt{r_{+}^{2}-r_{-}^{2}}}$
\begin{align}
r^{(Q)}_{-}>r_{+}.\label{eq137}
\end{align}
Thus, the horizons of the (QRN) are larger than those of the classical solutions.
\end{enumerate}

%%%%%%%%%%%%%%%%%%%%%%%%%%%%%%%%%%%%%%%%%%
\section{Discussion}

The primary purpose of this work was to construct a Schr\"{o}dinger like equation, instead of a Wheeler-DeWitt, for the Reissner-Nordström black hole, which holds the property of being covariant, under reparametrizations of the defined ``time'' variable. To do so, a different procedure was followed from those appearing in the literature, to the best of our knowledge. Specifically, the singular Lagrangian density \eqref{eq23}, \eqref{eq24}, \eqref{eq25} describing the dynamics of the system, was a functional of the degrees of freedom $(a,b,\grF)$ and the Lapse $n$. The variation of the minisuperspace action integral with respect to $n$ yield a constraint equation \eqref{eq21}, while dynamical equations appeared from the variations with respect to the other degrees of freedom \eqref{eq18}, \eqref{eq19}, \eqref{eq20}. Since many of the problems that appear in the quantization procedures are related to the existence of constraint equations, we thought that it is better if it did not exist, or better yet, if it were satisfied identically, once the solutions to the dynamical equations are obtained. One possible way to do so, without breaking the gauge freedom, is to solve the constraint equation with respect to the Lapse function. To this end, from the reduced equations, only two of them are independent \eqref{eq27}, \eqref{eq28}, while there are three degrees of freedom left $(a,b,\grF)$. We argued that this implies that the gauge freedom still exists and we could have arrived in these two independent equations from the beginning, if somehow we were able to guess the proper form of the Lapse as a function of the other degrees of freedom \eqref{eq26}.

The reduced Lagrangian density has a square root form \eqref{eq29} and is still singular if $(a,b,\grF)$ are all considered degrees of freedom. However, since there are only two independent equations, (the third is satisfied once the other two are solved) these equations of motion can be reproduced via the variation of the action from the reduced Lagrangian density, with respect to only two of the degrees of freedom. There is the freedom to choose any of the pairs $(a,b)$, $(a,\grF)$ or $(b,\grF)$. We have chosen $(a,\grF)$, hence the $b$ it is now considered as a mere function of the coordinate $r$ and not a degree of freedom. To this end, the Lagrangian density is regular with respect to $(a,\grF)$, it is ``time'' dependent due to the appearance of $b(r)$ and the gauge freedom still exists, encoded in the choice of $b(r)$. Note that due to the explicit position that $b(r)$ appears in the spacetime metric, we cannot gauge fix it to a constant value. 

As we have pointed out, the Lagrangian density has a square root form and as it is known, problems will arise in the quantization procedure when the square root Hamiltonian density will have to turned into an operator. To avoid such kind of problems, we have searched for a quadratic in the ``velocities'' Lagrangian density which reproduces the equations of motion and we succeeded in finding one \eqref{eq33}, \eqref{eq34},\,\eqref{eq35}. 

Before moving to the quantum description, we have noticed that the minisuperspace metric corresponds to a flat, Minkowski spacetime. This implies the existence of coordinates $(u,w)$ so that the minisuperspace metric acquires the diagonal form with eigenvalues $(-1,1)$. To this end, the reduced equations acquired a very simple form \eqref{eq42} and the solution in arbitrary gauge was obtained, by solving one algebraic and one linear second, order, differential equation. We find that this is one of the simplest ways to obtain the solution in arbitrary gauge.

At that point, a remarkable equivalence appeared: these equations of motion, have exactly the same form and are equal in number, with the equations of motion for a 3D electromagnetic pp-wave spacetime, that we have studied in some previous work \cite{Pailas:2019abb}. This bizarre and interesting coincidence intrigue us to assume that there might be a kind of classification of spacetimes+matter, based on the equivalence in form, of Einstein's equations, with the underlying reason been the geometrical characteristics of the minisuperspace metric. This idea, however, will be pursued in some future work.

To obtain the Hamiltonian density was an easy task. Note that since it is a ``time'' dependent quantity is not conserved. However, there is the $H_{con}$ \eqref{eq65} and three Noether charges \eqref{eq66} that are conserved modulo the equations of motion. For the quantum description we followed the canonical quantization procedure, meaning the elevation of observables into self-adjoint operators and the replacement of the Poisson brackets with the commutator. Since the Hamiltonian density is regular, the construction of a Schr\"{o}dinger like equation is possible. Furthermore, the hyperbolic nature of the Hamiltonian density operator implies that we have not defined an ``intrinsic time'' variable (we have not broken the gauge invariance). The idea is that we use as the parameter of ``time'' the coordinate (r), which appears explicitly in the Hamiltonian density through the function $M(r)$ (or $b(r)$ equivalently). This equation \eqref{eq75} looks like a hybrid between the ``intrinsic time'' Schr\"{o}dinger equation and the Wheeler-DeWitt equation that we described in the introduction. The difference with the first lies in the ``time''-covariance property, hence, the quantum descriptions for each different gauge choice are, by definition, equivalent. For the second, the difference appears in the evolution of states in our description, in contrast to the ``frozen'' picture related to the Wheeler-DeWitt equation. Note at this point that the word ``time'' was used nominally in the various places inside the text, since as it is inferred from the line element, the variable $r$ is a spatial coordinate. Alongside, any reference to evolution of states is with respect to this external parameter $r$ and should not be confused to the usual quantum mechanical sense of evolution of states in time.

The solution to this equation was obtained for wavefunctions that are common to the operators $\hat{H}_{con},\hat{Q}_{1},\hat{Q}_{2}$. The evolution of two initial states was studied, Dirac's delta and Gaussian. These two are approaching each other at a certain limit of the parameter that appears in the Gaussian distribution and control its width. For both cases, the wavefunctions were obtained and the probability densities were calculated \eqref{eq89} and \eqref{eq96} respectively. In order to acquire some sense of the quantum corrections to the spacetime and interpret the wavefunctions in geometrical terms, the Bohm analysis was employed. The results are the following: for the Dirac's delta initial state, the quantum potential is zero, hence the semi-classical ``trajectories'' coincide with the classical ones. The Gaussian initial state becomes a delta function only at a certain limit, hence the obtained result is a non-zero quantum potential, resulting different semi-classical from the classical ``trajectories'' \eqref{eq119}, \eqref{eq120}, \eqref{eq121}. The quantum corrections manifest themselves as the appearance of an additional energy-momentum tensor $T^{(Q)}_{\grm\grn}$ and a four-current density $J^{\grm}_{(Q)}$ in the system of Einstein's-Maxwell's equations. Their explicit formulas can be found in \eqref{eq124}-\eqref{eq127}. One important aspect of the semi-classical solutions is that they contain the classical solutions as a limit, specifically, in the limit that $T^{(Q)}_{\grm\grn}$, $J^{\grm}_{(Q)}$ vanish.

As we have pointed in the introduction, one of the motivations to construct a quantum theory of gravity, is for the purpose of ``healing'' the singularities from which the classical theory suffers. As we have obtained in the section IV.3, the quantum corrected RN is still singular. One way to observe this, is due to the Ricci scalar, whose appearance is solely due to $T^{(Q)}_{\grm\grn}$. For small values of the variable $r$, this objects scales as $R\sim \frac{1}{r^{3}}$, hence as $r\rightarrow 0$ the Ricci scalar tends to infinity. At that point, since for the classical RN solution the Ricci scalar is zero, we found more insightful to compare the Kretschmann scalars of these two spacetimes. We have defined a criterion in order to adjudicate whether or not the situation gets better in the case of QRN regarding the divergence of the Kretschmann scalar. The criterion states the following: 

\textit{Between two spacetimes which suffer from the same kind of singularity, in this specific case a point-like, ``less'' singular will be characterized the one whose singularity is more localized around the point of interest.}

Regarding the QRN and RN black holes, there are admissible values of the parameters for which QRN is ``less'' singular, as well as values for which is ``more'' singular. In the figures that are presented, there is a tendency which states that: as $\grb\rightarrow \infty$, (translated as stronger quantum effects, moving further away from Dirac's delta distribution) the characteristics of these black holes which are encoded as values of $\gra$ for which QRN is still ``less'' singular are closing to zero. The upper value of $\gra$ for a specific value of $\grb$, depends also on the proximity to the point $r=0$. 

Another interesting result also related to the singularities is the following: Since we have obtained a well defined Hilbert space and we have calculated the probability densities, we may follow DeWitt's proposal \cite{PhysRev.160.1113} which states that: if the wave function or better the probability density vanishes at the configuration points where the classical singularity appears, then it is avoided at the quantum level. The classical singularity appears for $r\rightarrow 0$ which translates into $\grt\rightarrow \infty,\,\grs\rightarrow \infty$. By carefully evaluating at this limit, the probability densities for both cases of initial states, \eqref{eq89}, \eqref{eq96} we find that they are equal to zero. Hence, in the light of the above proposal, the singularities disappear in the quantum regime. Therefore, a discrepancy appears between this result and the result obtained via Bohm's analysis. One possible explanation it might be that Bohm's analysis is just an approximate scheme, with the approximation localized in the identification \eqref{eq103} of the classical and quantum momenta. A recent article discussing the singularity resolution and its dependence on the choice of clock  can be found in \cite{Gielen:2020abd}.

Since we have used the probability densities to comment on the existence of singularities, it is rather instructive to comment also on the possible interpretation of these probability densities. As we have explained, the external parameter $r$ is considered as the ``time'' variable in the Schr\"{o}dinger equation. To this end, the $P(r,\grt,\grs)=|\grC(r,\grt,\grs)|^{2}$ is to be understood as the probability density for the variables $(\grt,\grs)$ at ``time'' $r$. This kind of interpretation is not unknown, it appears also in the ``intrinsic time'' formalism of the full quantum theory, where a multi- ``time'' Schr\"{o}dinger equation is constructed. For more information \cite{Isham:1992ms}. In this particular work, we have used symmetries to restrict ourselves in a minisuperspace model, which implies, due to the minisuperspace model at hand, that a spacelike parameter $r$ remains to be interpreted as ``time''. If a cosmological minisuperspace model was to be used instead, then the usual timelike coordinate $t$ would appear. Regarding the collapse of the wave-function after some measurement we may say that the usual quantum mechanical interpretation could be used. Lastly, in the lack of a complete physical theory of quantum gravity, it is far from obvious to know whether such a quantum minisuperspace model would appear as a limiting case of the full theory and hence render it physical. At this point in time, we would say that it is more of a mathematical model with a possible physical meaning through the use of the Bohmian interpretation; it might be possible to contrast the quantum corrected metric with observations.

Event horizons (inner and outer) also appear for QRN, under the assumption that the constant related to the quantum effects, $l_{p}$, satisfies the condition $l_{p}>r_{+}$, where $r_{+}$ the outer RN horizon. In comparison with the event horizons of the classical solution we have found that they have a larger radius. The reason for that may be attributed as well to the presence of the additional energy-momentum tensor in the quantum corrected metric. There exist the case of extremal horizon as well. Also, for different values of the parameter $l_{p}$, which controls the localization of $T^{(Q)}_{\grm\grn}$ and $J^{\grm}_{(Q)}$, there are the cases where two horizons exist, only one degenerate and none, implying a naked singularity. We have observed that with the above defined criteria for the singularity resolution, the naked singularity does not disappear. Of course, let us point out once again, that based on DeWitt's proposal there are no singularities at all. Hence, this is another strand of the revealed discrepancy. All the results that have been obtained via Bohm's analysis, can be reduced to the Schwarzschild black hole by the simple substitution $r_{q}=0$. For future work, it may be interesting to study the geodesic equation and find the corrections that are imposed due to quantum effects. Furthermore, some experimental boundaries may be imposed to the values of the parameter $l_{p}$.

Finally, we believe that this ``hybrid'' procedure on the construction of a Schr\"{o}dinger equation, for the case of minisuperspace models, will provide some new perspective on the problem of time. It is our intention to study, in some future work, other minisuperspace models as well and perhaps, even try to formulate this procedure in the full theory of gravity.

%%%%%%%%%%%%%%%%%%%%%%%%%%%%%%%%%%%%%%%%%%
%%%%%%%%%%%%%%%%%%%%%%%%%%%%%%%%%%%%%%%%%%
\section{Conclusions}

Since the Discussion section appears quite lengthy, we find helpful to present in short the important points.
\begin{enumerate}
\item Reduction of the singular minisuperspace Lagrangian (Hamiltonian) density of the Reissner-Nordstr\"{o}m black hole, to a regular, ``time''-dependent Lagrangian (Hamiltonian) density, without gauge fixing.
\item Equivalence revealed at the level of equations, between the above described spacetime and a 3D electromagnetic pp-wave studied in previous work. 
\item Construction of a ``time''-covariant Schr\"{o}dinger equation, as a hybrid between the ``intrinsic'' time Schr\"{o}dinger and Wheeler-DeWitt equation.
\item Singularity criterion: ``less'' singular between two spacetimes is characterized the one whose singularity is more localized around the point of interest.
\item Discrepancy revealed concerning the existence of singularity, between the geometric in origin, Bohm's analysis and DeWitt's proposal (zero probability density at the classical singular point).
\end{enumerate}
%%%%%%%%%%%%%%%%%%%%%%%%%%%%%%%%%%%%%%%%%%
\section{Materials and Methods}

The \textit{Mathematica} \textcopyright\, software was used wherever was needed.

%%%%%%%%%%%%%%%%%%%%%%%%%%%%%%%%%%%%%%%%%%
\vspace{6pt} 

%%%%%%%%%%%%%%%%%%%%%%%%%%%%%%%%%%%%%%%%%%
%% optional
%\supplementary{The following are available online at \linksupplementary{s1}, Figure S1: title, Table S1: title, Video S1: title.}

% Only for the journal Methods and Protocols:
% If you wish to submit a video article, please do so with any other supplementary material.
% \supplementary{The following are available at \linksupplementary{s1}, Figure S1: title, Table S1: title, Video S1: title. A supporting video article is available at doi: link.}

%%%%%%%%%%%%%%%%%%%%%%%%%%%%%%%%%%%%%%%%%%

%%%%%%%%%%%%%%%%%%%%%%%%%%%%%%%%%%%%%%%%%%

%%%%%%%%%%%%%%%%%%%%%%%%%%%%%%%%%%%%%%%%%%

\section{Acknowledgments}
I wish to express my warmest gratitude to my supervisor, Professor Theodosios Christodoulakis, for patiently reading the manuscript and offering his advice.

\bibliographystyle{unsrt}

\end{document}